\documentclass[structabstract]{aa}  
\usepackage{graphicx}
\usepackage{natbib}
\usepackage{txfonts}
\begin{document}
   \title{Stellar activity of planetary host star HD 189733\thanks{ Based on observations collected with the {\it SOPHIE}  spectrograph on the 1.93-m telescope at Observatoire de Haute-Provence (CNRS), France, by the {\it SOPHIE}  Consortium (program 07A.PNP.CONS)}}

   \author{I. Boisse
          \inst{1}
          \and
          C. Moutou\inst{2}
          \and
          A. Vidal-Madjar\inst{1}
          \and
          F. Bouchy\inst{1}
          \and
          F. Pont\inst{3} 
          \and
          G. H\'ebrard\inst{1}
          \and
          X. Bonfils\inst{5}\fnmsep\inst{6}
           \and
          B. Croll\inst{7} 
          \and
          X. Delfosse\inst{4}
          \and
          M. Desort\inst{4}
          \and
          T. Forveille\inst{4}
          \and
          A.-M. Lagrange\inst{4}
          \and
          B. Loeillet\inst{2}
          \and
          C. Lovis\inst{5}
          \and
          J. M. Matthews\inst{8}
          \and
          M. Mayor\inst{5}
          \and
          F. Pepe\inst{5}
          \and
          C. Perrier\inst{4}
          \and
          D. Queloz \inst{5}
          \and
          J. F. Rowe\inst{8}
          \and
          N. C. Santos\inst{6}
          \and
          D. S\'egransan\inst{5}
          \and
          S. Udry\inst{5}      
          }

   \institute{Institut d'Astrophysique de Paris, CNRS (UMR 7095)-Universit\'e Pierre \& Marie Curie, 98bis bd. Arago, 75014 Paris, France
              \email{iboisse@iap.fr}
        \and
            Laboratoire d'Astrophysique de Marseille, CNRS (UMR 6110)-Universit\'e de Provence, P\^ole de l'Etoile Site de Ch\^ateau-Gombert,
38 rue Fr\'ed\'eric Joliot-Curie, 13388 Marseille cedex 13, France 
         \and
           School of Physics, University of Exeter, Stocker Road, Exeter EX4 4QL, United Kingdom 
         \and
            Laboratoire d'Astrophysique de Grenoble, CNRS (UMR 5571)-Universit\'e Joseph Fourier, BP 53, 38041 Grenoble Cedex 9, France
         \and 
            Observatoire de Gen\`eve, Universit\'e de Gen\`eve, 51 Ch. des Maillettes, 1290 Sauverny, Switzerland
       \and 
         Centro de Astrof{\'\i}sica, Universidade do Porto, Rua das Estrelas, 4150-762 Porto, Portugal
         \and
         Dept. of Astronomy \& Astrophysics, Univ. Toronto, 50 George St., Toronto, ON M5S 3H4, Canada
         \and
         University of British Columbia, 6224 Agricultural Road, Vancouver, BC V6T 1Z1, Canada
             }

   \date{Received date; accepted date}

% \abstract{}{}{}{}{} 
% 5 {} token are mandatory
 
  \abstract
   {}
 {Extra-solar planet search programs require high-precision velocity measurements. They need to determine how to differentiate between radial-velocity variations
due to Doppler motion and the noise induced by stellar activity.}
   {We monitored the active K2V star HD~189733 and its transiting planetary companion, which has a 2.2-day orbital period. We used the high-resolution spectograph \textit{SOPHIE}  mounted on the 1.93-m telescope at the Observatoire de
Haute-Provence to obtain 55 spectra of HD~189733 over nearly two months. We refined the HD~189733b orbit parameters and placed limits on both the eccentricity and long-term velocity gradient. After subtracting the orbital motion of the planet, 
we compared the variability in spectroscopic activity 
indices with the evolution in the radial-velocity residuals and the shape of spectral lines.}
   {The radial velocity, the spectral-line profile, and the activity indices measured in \ion{He}{I}  (5875.62~$\AA$),
   H$\alpha$ (6562.81~$\AA$), and both of the \ion{Ca}{II}~H\&K lines (3968.47~\AA\ and 3933.66~\AA, respectively) exhibit a periodicity close to the stellar-rotation period and the 
correlations between them are consistent with a spotted stellar surface in rotation. We used these correlations to correct for the radial-velocity jitter due to stellar activity. This results in achieving high precision in measuring the orbital parameters, with a semi-amplitude $K=200.56 \pm 0.88$ m$\cdot$s$^{-1}$ and a derived planet mass of $M_{P}=1.13 \pm 0.03$ M$_{Jup}$.}
   {}


   \maketitle
%
%________________________________________________________________

\section{Introduction}
   The precision of radial-velocity (RV) searches for exoplanets depends on the noise induced by photospheric luminosity variations due to
   active regions (e.g. Santos et al. 2002). 
   Until now, most active stars have been rejected from major RV surveys, but these systems might be interesting objects such as HD~189733.
   Planets around active stars are now found by research instruments such as
   CoRoT, and need radial-velocity follow-up \citep{alonso}. 
   Monitoring the activity
   of active planetary-host stars may provide a means of disentangling radial-velocity variations
   due to Doppler motion from the noise induced by stellar activity, i.e. dark spots in the photospheric region.  
   
   HD~189733 is a well-known planet-hosting system discovered by Bouchy et al. (2005) in a metallicity-biased search for
   transiting hot Jupiters with the ELODIE spectrograph, then mounted on the 1.93m telescope at Observatoire de Haute-Provence (OHP). It is one of the two brightest transiting systems (V~=~7.7) and thus the parameters of the system are well constrained, thanks to a series of
   detailed spectroscopic and photometric monitoring programs \citep{bouchy,bakos,Winn07}. Such monitoring constrains the orientation of the system and the possibility of there being
   another planet in the system \citep{croll,miller}. For instance, $HST$ photometry 
   excluded with high probability  the presence of Earth-sized moons and Saturn-type debris rings around the planet, and
   excluded a massive second planet in outer 2:1 resonance (Pont et al. 2007).
   
   HD~189733 is a chromospherically active star. This dwarf star is classified as a variable of type BY Dra, i.e. K stars with quasiperiodic photometric variations, on timescales ranging from hours to years. H\'ebrard \& Lecavelier des Etangs (2006) detected stellar microvariability in
  \textit{Hipparcos} photometric data. Winn et al. (2006) found
  that HD~189733 exhibits velocity noise or "photospheric jitter" with an amplitude of 12~m s$^{-1}$.
  Henry \& Winn (2008) used the optical photometric variability due to spots and plages carried into
  and out of view as the star rotates to determine the rotation period of the star.  
  
  Activity is a signature of stellar magnetism. Moutou et al. (2007) completed spectropolarimetric measurements of
  HD~189733 to derive theoretical models of planet-star magnetic interaction with quantitative constraints. They observed activity signatures that varyied mostly with the
  rotation period of the star. They found that the chromospheric-activity index measured in the \ion{Ca}{II}~H\&K lines and Zeeman signatures are
  consistent with a moderate field being present at the stellar surface that is modulated by the rotation cycle. Shkolnik et al. (2007) monitored the \ion {Ca} {II}~H\&K lines chromospheric index and associated the index variability with an active region rotating with the star. Moreover, they identified an additional flare 
potentially induced by interactions with the hot Jupiter.
   
   In this article, we monitor activity using all the parameters available from high-precision spectroscopic measurements: radial-velocity, bisector of the lines, and chromospheric indices measured in \ion {He} {I}  (5875.62~\AA),
   H$\alpha$ (6562.81~\AA), and the \ion {Ca} {II}~H\&K lines (3968.47~\AA\ and 3933.66~\AA, respectively).
   
   Section~2 describes the observations and the data reduction process. The refined parameters of the HD~189733b orbital solution are discussed in Sect.~3. In Section~4, we examine the radial-velocity residuals
    and the bisector behavior, followed by a study of the chromospheric activity in Sect.~5. In the last section, we analyze the correlation between parameters and how this analysis may be used to disentangle radial-velocity variation from the noise induced by activity-related phenomena.

%__________________________________________________________________

\section{Observations and data reduction}

Fifty-five radial-velocity (RV) measurements were obtained between 12 July 2007 and 23 August 2007 with the high-resolution {\it SOPHIE}  spectrograph \citep{bouchy06}
 mounted on the 1.93-m telescope at the Observatoire de Haute-Provence, France.  We gathered about
 two exposures per night to sample correctly both the short orbital period  ($\simeq$~2.2~days) and the longer rotation period of the star  ($\simeq$~12~days). This {\it SOPHIE}  campaign was performed simultaneously with a $MOST$ photometric campaign, discussed shortly hereafter.

The observations were conducted in the high-resolution mode ($R$=$\lambda/\delta\lambda$ $\sim$75\,000). The simultaneous Thorium-Argon (ThAr) lamp was used to achieve higher
precision in RV measurements.
The typical exposure
time was 6 minutes, sufficiently long to reach an average signal-to-noise ratio (SNR) per pixel of about 80 at $\lambda$ = 5500~\AA. RVs  were determined using a weighted cross-correlation method with a numerical G2 mask: this mask was 
constructed from the Sun spectrum atlas including up to 3645 lines, following the procedure of Baranne et al. (1996) and {Pepe et al. (2002)} implemented in the automated reduction package provided for {\it SOPHIE}. We removed the first 8 blue spectral orders containing essentially noise
and the last 3 orders in the red part that distorted the RV calculation due to the presence of telluric lines. {The effective wavelength range used for the RV calculation was between 4311 \AA\ and 6551 \AA.}

The typical photon-noise error of the measurements was about $1.5$ m s$^{-1}$ and the mean uncertainty in the wavelength calibration was around $1.0$ m s$^{-1}$. The main limitation of {\it SOPHIE}  originates in the Cassegrain Fiber Adaper  (developed for the previous ELODIE 
spectrograph), which is responsible for the guiding and centering system in the fiber entrance. 
Empirically, the stable stars followed by the {\it SOPHIE}  Consortium have an accuracy around $3$ m s$^{-1}$ so we added 
a systematic error of $3$ m s$^{-1}$. Our average error was then around $3.5$ m s$^{-1}$, which was the quadratical sum of the photon-noise error, the uncertainty in the wavelength calibration, and an estimation of the {\it SOPHIE}  current systematics.

A perturbation in the apparent radial velocities of the star is caused by the planet blocking part of the Doppler-shifted light from the rotating star. The resulting radial velocity anomaly due to the transit of the planet in front of the star is called the Rossiter-McLaughlin effect.
Five measurements were made during the transit time, calculated with ephemerides from Winn et al. (2007). We removed them from the study, plus another measurement of too low SNR ($ \leqslant$ 20 at $\lambda = 550$ nm) due to bad weather.

Background Moon light also polluted part of the exposures, shifting the RV measurements by up to $40$ m s$^{-1}$. Indeed, the sky acceptance of the {\it SOPHIE}  fiber is 3 arc-second and the spectra are contaminated by sky background. Clouds or full Moon on barycentric Julian date BJD-2454000\ =\ 311 days increases the level of polluting light. The Moon light shifts the measured RVs significantly because the Moon apparent velocity changes from $+7$ km s$^{-1}$ to $-9$ km s$^{-1}$ between 12 July 2007 and 23 August 2007, whereas the mean RV of the star is around $ -2.3$ km s$^{-1}$. The sky background may be recorded using the second fiber feeding the {\it SOPHIE}  spectrograph but in our case, this fiber was used by the ThAr calibration lamp. Consequently, we removed 16 measurements of RVs and bisectors when there was a significant correlation peak  with RV $\simeq$ Moon apparent velocity in the second fiber during an exposure of another star in less than two hours of the HD189733 exposures. These spectra were not removed from the activity studies in Sect. 5 because the Moon light did not pollute strongly the flux in the lines used as activity indicators. Finally, 33 measurements of RVs and bisector velocity span were studied. These measurements are available in the version of Table 1 at the CDS, which contains in its cols. 1, 2, and 3, the time in BJD, the RV, and the error in RV, respectively.
 % Table 1 available electronically only
\onltab{1}{
\begin{table*}%t2
\caption{HD 189733 RV measurements}\label{Properties}
\begin{tabular}{ccc}
            \hline
            \noalign{\smallskip}
             BJD  [day] & RV [km.$^{-1}$] & RV error [km.$^{-1}$] \\
            \noalign{\smallskip}
            \hline
            \noalign{\smallskip}           

2454294.5084 & -2.0865 & 0.0033  \\
2454298.3704 & -2.2320 & 0.0037 \\
2454298.6022 & -2.1180 & 0.0034 \\
2454299.6075 & -2.3931 & 0.0034 \\
2454300.3518 & -2.3702 & 0.0034 \\
2454300.6136 & -2.2123 & 0.0035 \\
2454301.3613 & -2.1437 & 0.0034 \\
2454302.3864 & -2.4464 & 0.0035 \\
2454303.6148 & -2.1555 & 0.0035 \\
2454304.3853 & -2.4705 & 0.0035 \\
2454306.3676 & -2.4206 & 0.0038 \\
2454306.5803 & -2.4638 & 0.0036 \\
2454306.6077 & -2.4644 & 0.0033 \\
2454307.3669 & -2.1593 & 0.0034 \\
2454307.4946 & -2.1057 & 0.0033 \\
2454307.5826 & -2.0795 & 0.0033 \\
2454308.3698 & -2.3264 & 0.0033 \\
2454308.5947 & -2.4465 & 0.0033 \\
2454309.4000 & -2.2859 & 0.0034 \\
2454309.5121 & -2.2181 & 0.0033 \\
2454315.5672 & -2.4759 & 0.0038 \\
2454316.3922 & -2.1139 & 0.0034 \\
2454316.5673 & -2.0794 & 0.0034 \\
2454317.4941 & -2.4306 & 0.0034 \\
2454317.6281 & -2.4652 & 0.0034 \\
2454318.3294 & -2.2216 & 0.0036 \\
2454318.5691 & -2.1097 & 0.0034 \\
2454328.3935 & -2.3662 & 0.0033 \\
2454328.5365 & -2.4255 & 0.0042 \\
2454330.3345 & -2.1936 & 0.0037 \\
2454334.3945 & -2.0675 & 0.0034 \\
2454335.3983 & -2.4625 & 0.0035 \\
2454336.4651 & -2.0804 & 0.0034 \\
 \noalign{\smallskip}
  \hline

\end{tabular}
\end{table*}
}% end of onltab

%__________________________________________________________________

\section{Determination of system parameters}
%%
%                                                One column figure
%----------------------------------------------------------- S_vib
   \begin{figure}
   \centering
    \includegraphics[width=9cm]{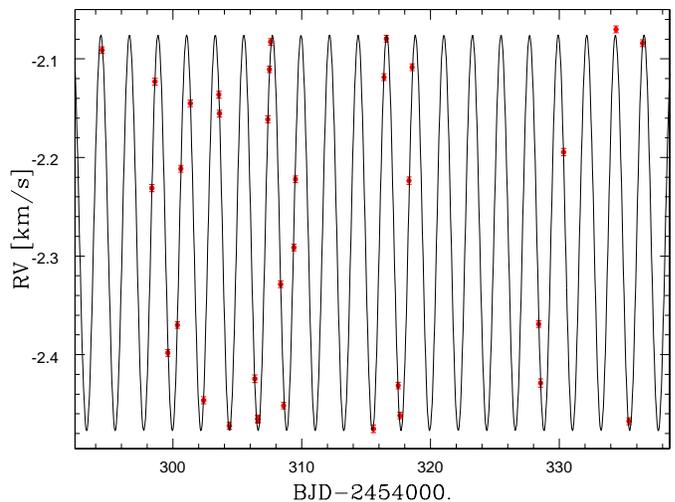}
      \caption{ Radial velocities of HD 189733 derived from {\it SOPHIE}  spectra collected in July and August 2007 as a function of
time. The refined Keplerian orbital solution is plotted as a solid
line. Individual error bars are also plotted.}
         \label{myfig0}
   \end{figure}
%
%______________________________________________________________

\begin{figure}
   \centering
   \includegraphics[width=9cm]{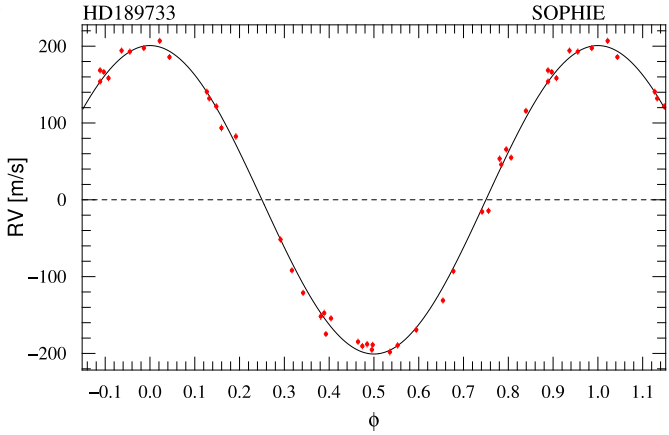}
      \caption{Radial velocities of HD 189733 derived from {\it SOPHIE}  spectra collected in July and August 2007 as a function of
orbital phase. The refined Keplerian orbital solution is plotted as a solid
line. Individual error bars are also plotted.
              }
         \label{myfig1}
   \end{figure}

The Keplerian solution was derived for the 33 RV points selected in Sect.2. We fixed the orbital period and the time of periastron, well constrained by previous studies, using transit timing  \citep{pont,croll,Winn07}. We used the values estimated by Winn et al. (2007), $P=2.2185733 \pm 0.0000019$~days and our {periastron time} $T_{0}$ was derived from their time of transit center $T_{T}=2453988.8034 \pm 0.0002$~[BJD]. {\it SOPHIE} RV measurements were then most accurately fitted with a Keplerian function of semi-amplitude 
$K=200.6 \pm 2.3$ m s$^{-1}$, an eccentricity consistent with zero and a mean radial velocity ${V}=-2.277$~km s$^{-1}$ (Figs.~\ref{myfig0} and ~\ref{myfig1}, Table 2), in agreement with previous results \citep{bouchy,Winn}.  Our determination of the semi-amplitude is more accurate than the value reported by Bouchy et al. (2005) and the value evaluated from the measurements published by Winn et al. (2006).

%
%                                                One column figure
%----------------------------------------------------------- S_vib
%
%______________________________________________________________

A zero value for the eccentricity is satisfactory because one would expect the orbit of the hot Jupiter HD~189733b to be nearly circular due to tidal effects. Nevertheless, Knutson et al. (2007) discussed the possibility of a weak eccentricity $e \cos \omega= 0.0010\ \pm\ 0.0002$, where $\omega$ is the longitude of pericenter for HD~189733b, to explain a time delay of 150 $\pm$ 24 s in the photometric observations of the secondary transit.  Similarly, Winn et al. (2007) determined the limits the orbital eccentricity and found that eccentricity might be as high as $e$ $\approx$ 0.02 if $\omega \approx \pm$ 90$^{\degr}$.   

We assessed for the detectability of non-zero eccentricity for the orbit of HD~189733b in the RV data set. We used all published RV measurements, a total of 151 values, \citep{bouchy,Winn07} in addition to our own, excluding all measurements for data acquired during the transit. 
We fitted the data with a long-term linear-velocity gradient $v$, taking into account that HD~189733 has an M dwarf stellar companion (Bakos et al. 2006a) with an orbital period of $\sim$ 3200 years and an RV amplitude of $\sim$ 2 km s$^{-1}$. The trend identified between the three datasets was $v$ =1.4 $\pm$ 2.3 m~s$^{-1}$~yr$^{-1}$. This is consistent with the presence of the stellar companion and does not require a long-period planetary companion to be explained. 
Leaving $e$ and $\omega$ as free parameters, we determined that the eccentricity was lower than~0.008. The long-term velocity gradient does not have a significant effect on the determination of the eccentricity. The result agrees with the value of Winn et al. (2007) and is still consistent with the Knutson et al. (2007) solution introduced to explain the secondary transit delay.                         
   
%__________________________________________________ One column table
   \begin{table}
      \caption{HD 189733 system parameters}
        \begin{tabular}{l l}
            \hline
              Parameter & Value\\
            \hline
             P$_{rot}$ [days] & 11.953 $\pm$ 0.009$^{\mathrm{a}}$  \\
             M$_{*}$ [M$_{\sun}$] & 0.82 $\pm$  0.03$^{\mathrm{b}}$  \\
             P$_{p}$ [days] & 2.2185733\  [fixed] $^{\mathrm{c}}$\\
            T$_{0}$ [BJD]  & 2453988.24876\ [fixed]  $^{\mathrm{c}}$\\
            T$_{T}$ [BJD]  & 2453988.8034\ [fixed]  $^{\mathrm{c}}$\\
            v $\sin$ i$_{\ast}$ [km s$^{-1}$] & 2.97$\pm$ 0.22  $^{\mathrm{c}}$\\
             i [$^{\degr}$]  & 85.68 $\pm$ 0.04 $^{\mathrm{d}}$\\
            e  & $\leqslant$ 0.008 \\
            V [km s$^{-1}$] & -2.2765 $\pm$ 0.0017 \\
            K [m s$^{-1}$]\ (before\ correction) & 200.60 $\pm$ 2.32 \\
            K [m s$^{-1}$]\ (after\ correction) & 200.56 $\pm$ 0.88 \\
            M$_{p}$ [M$_{Jup}$]  & 1.13 $\pm$ 0.03  \\
            \hline
         \end{tabular}
         
\begin{list}{}{}
\item[$^{\mathrm{a}}$] Henry \& Winn (2008)
\item[$^{\mathrm{b}}$] Bouchy et al. (2005)
\item[$^{\mathrm{c}}$] Winn et al. (2007)
\item[$^{\mathrm{d}}$] Pont et al. (2007)
\end{list}
   \end{table}
%

%__________________________________________________________________
   
%_____________________________________________________________

\section{RV residuals and bisectors of the lines}

  The large "observed-minus-calculated" (O-C) residuals around the best-fit Keplerian solution are usually explained by activity-induced "jitter" of the star. Our dispersion value, $9.1$ m s$^{-1}$, is smaller than the $12$ m s$^{-1}$ measured by Winn et al. (2007) and the $15$~m s$^{-1}$ residuals found by Bouchy et al. (2005). These larger residuals might be due to the amplitude of the stellar-activity jitter decreasing with time (within long-term stellar-activity cycle) and to the lower accuracy of ELODIE spectrograph used by Bouchy et al. (2005).

%
%                                                One column figure
%----------------------------------------------------------- S_vib
   \begin{figure}[b]
   \centering
   \includegraphics[width=9cm]{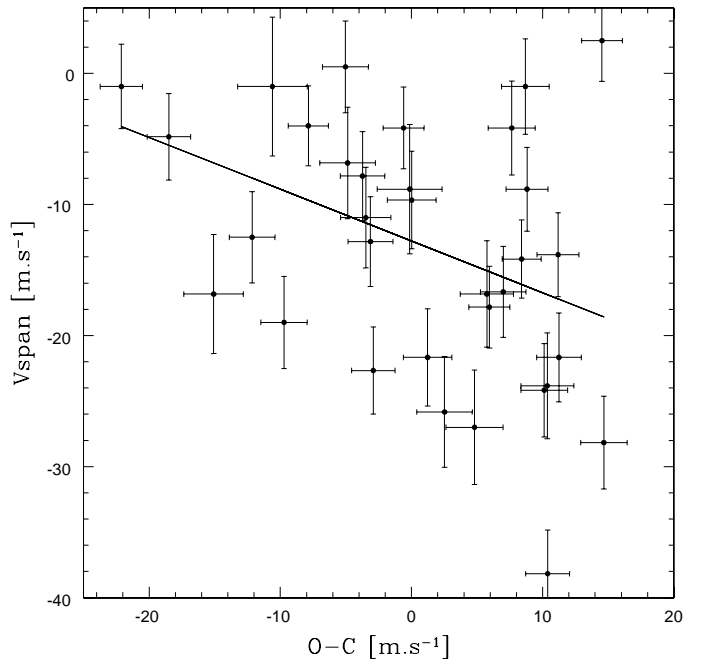}
      \caption{The residuals (O-C) from the Keplerian fit versus bisector span of the CCF profile. We observe an anticorrelation, which is expected in case of RV variations induced by stellar activity. The probability that the 34 observations come from a random sample of two uncorrelated variables is lower than 0.1 \%.
              }
         \label{myfig5}
   \end{figure}

The cross-correlation function (CCF), calculated for measuring the RVs of the star, corresponds to an average of all the spectral lines. We analyzed the CCF bisector velocity span (V$_{span}$), a measurement of the shape of the CCF as defined in Queloz et al. (2001). Active regions at the stellar surface, such as dark spots or bright plages, induce an asymmetry in the emitted light of the star changing periodically as they move in and out of view with stellar rotation. They then also induce an asymmetry in the line profile that can be measured by the V$_{span}$.
An activity phenomenon
is characterized by an anticorrelation between the observed radial-velocity (here, the residuals from the fit) and V$_{span}$ \cite{queloz}. In Figure.~\ref{myfig5}, the V$_{span}$ is anticorrelated with the residuals in agreement with a stellar phenomenon as an explanation of the RV variations. V$_{span}$ error bars correspond to twice the size of the RV error bars. The calculation of the V$_{span}$ needs two measurements of RV, which provides an approximate estimate of its uncertainty. The linear correlation coefficient is -0.58, implying that the null-hypothesis, i.e. that the 34 observations come from a random sample of two uncorrelated variables, has a probability of lower than 0.1 \%. A linear fit to the relation of V$_{span}$ versus residuals (O-C) is given by V$_{span}$ $ = (-0.61 \pm 0.15) $(O-C)$-(12.6\pm 1.4)$.

\section{Spectroscopic indices of stellar activity}

\subsection{\ion {Ca} {II}~H\&K lines}
 
%                                                One column figure
%----------------------------------------------------------- S_vib
   \begin{figure}[b]
   \centering
  \includegraphics[width=9cm]{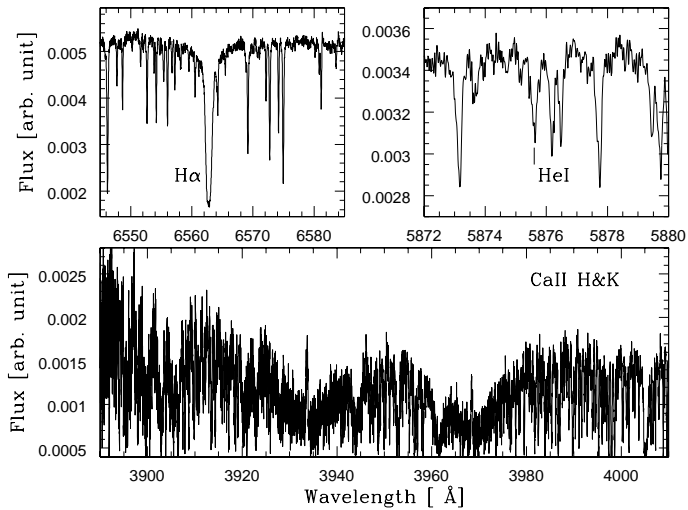}
       \caption{H$\alpha$ (top left) and \ion {He} {I}  (top right) and emission reversal in the \ion {Ca} {II}~H and K lines of the star HD 189733 (bottom) in the {\it SOPHIE} spectrograph typical spectrum. 
              }
         \label{myfig6}
   \end{figure}

The \ion {Ca} {II}~H and K lines are broad, deep lines for solar-type stars. Respectively centered on 3968.47~\AA \ and 3933.66~\AA,
 (Fig~\ref{myfig6}), the emission reversal in the core of the \ion {Ca} {II}~H$\&$K resonant lines, resulting from non-radiative heating of the
chromosphere, is indicative of spots and plages, i.e. active regions.
{When studying the \ion {Ca} {II}~H$\&$K lines in SOPHIE spectra, one must account for the fact that they are located in a spectral region that suffers from contamination from the simultaneously recorded ThAr spectrum.}
The star spectrum is recorded in fiber A, while the ThAr lamp calibration spectrum is recorded simultaneously in fiber B. The calibration lamp produces a background of scattered
light all over the CCD at the level of $0.1\%$ of the stellar continuum. In order to use the signature of flux emitted in the \ion {Ca} {II}  lines, we substract this background level from each stellar spectrum. We define our \ion {Ca} {II}~H+K index following the Mt. Wilson S index \citep{Wilson,baliunas}:
\begin{equation}
\centering
Index=\frac{H+K}{B+V}
\end{equation}
 
 \noindent where H and K are the flux measured in a 1.09 \AA\ wide window centered on each line of the \ion {Ca} {II} doublet, and B and V
 estimate the continuum on both sides with the flux measured in 20~\AA\ wide windows respectively centered on 3900~\AA\ and
 4000~\AA. We determined the errors of the \ion {Ca} {II} index by taking into account the CCD readout and photon shot noise. We were unable to estimate the error due to the background substraction because it depends on too many parameters: SNR, exposure time, elevation of the target, and the weather. So, this index is mainly dominated by systematics.

\begin{figure}
   \centering
   \includegraphics[width=9cm]{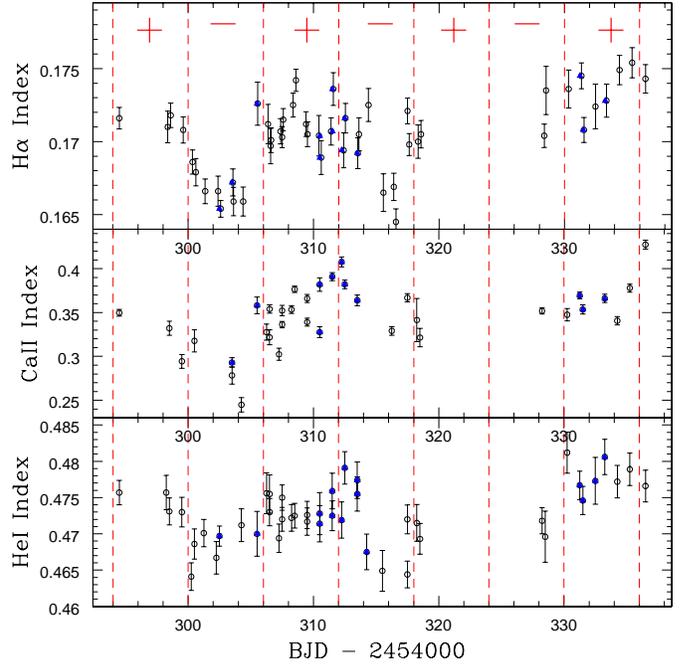}
      \caption{Evolution of our H$\alpha$ (top), \ion {Ca} {II} (middle) and \ion {He} {I}  (bottom) activity indices as a function of time. Error bars are also plotted. The red dashed lines indicate roughly the periods of minimal and maximal stellar activity based on the H$\alpha$ index and separated by half the rotation period of the star. The blue filled points come from spectra rejected in the RVs and V$_{span}$ studies (see text).
                    }
         \label{myfig7}
   \end{figure}

We removed from this study all spectra for which the \ion {Ca} {II} spectral range had insufficient flux due chiefly
to the atmospheric refraction that dispersed the blue part of the spectra more significantly. We plot the calculated index for the remaining 38 spectra  as a function of time in Fig.~\ref{myfig7} (middle panel) and observe a decrease in the activity level around $BJD-2454000\ \cong 304$ days.
%
%______________________________________________________________
 %
%                                                One column figure
%----------------------------------------------------------- S_vib
%
%______________________________________________________________

\subsection{H$\alpha$ line}

The H$\alpha$ line at $\lambda = 6562.808~\AA$ (Fig.~\ref{myfig6}) is similarly sensitive to the mean chromospheric activity \citep{kurster,bonfils}. H$\alpha$ is less affected by the flux of the ThAr calibration lamp than \ion {Ca} {II} lines. We defined our index as described in Bonfils et al. (2007):
\begin{equation}
\centering
Index=\frac{F_{H\alpha}}{F_{1}+F_{2}}
\end{equation}
with $F_{H\alpha}$ sampling the H$\alpha$ line and $F_{1}$ and $F_{2}$ the continuum on both sides of the line. The $F_{H\alpha}$ interval is 0.678 \AA\ wide and centered on $6562.808~\AA$, while $F_{1}$ and $F_{2}$ are integrated over $[6545.495-6556.245]~\AA$ and $[6575.934-6584.684]~\AA$, respectively. The errors bars are also calculated in a similar way.

We used all 55 available spectra. Figure~\ref{myfig7} shows the variations in the H$\alpha$ index with time; it is characterised by a lower noise level than the \ion {Ca} {II} index, as expected, and by two dips in the activity level around $BJD-2454000 \cong$ 303 and 316 days.

 \subsection{\ion {He} {I} line}
 
 The absorption line of the \ion {He} {I} D3 triplet at $\lambda = 5875.62~\AA$ (Fig.~\ref{myfig6}) is a signature of a non-radiative heating processes. The \ion {He} {I} is known to be a tracer of the solar plages (Landman 1981). Since it is sensitive to the heating of the chromosphere (Saar et al. 1997), this absorption line is also a diagnostic of stellar magnetic activity. We define the \ion {He} {I} index in a similar way to that for H$\alpha$ :
 \begin{equation}
\centering
Index=\frac{F_{\ion {He} {I} }}{F_{1}+F_{2}}
\end{equation}
with $F_{\ion {He} {I} }$ sampling the \ion {He} {I} line, and $F_{1}$ and $F_{2}$ representing the continuum on both sides of the line. Our $F_{He I}$ is centered on $\lambda = 5875.62~\AA$ and is $0.4\ \AA$ wide, while $F_{1}$ and $F_{2}$ are integrated over $[5874.0-5875.0]~\AA$ and $[5878.5-5879.5]~\AA$, respectively. 
The \ion {He} {I} line is located at the borders of two spectral orders in our SOPHIE spectra and is far noisier than the H$\alpha$ line. Moreover, in this region of the spectra, telluric lines increase the noise level. Nevertheless, two decreases of the index around $BJD-2454000 \cong$ 302 and 315 days are evident in Fig.~\ref{myfig7}, and {perhaps a partial third} one around 328 days. The behavior of this index reinforces the active region interpretation.\\

%                                                One column figure
%----------------------------------------------------------- S_vib
   \begin{figure}[t]
   \centering
      \includegraphics[width=9cm]{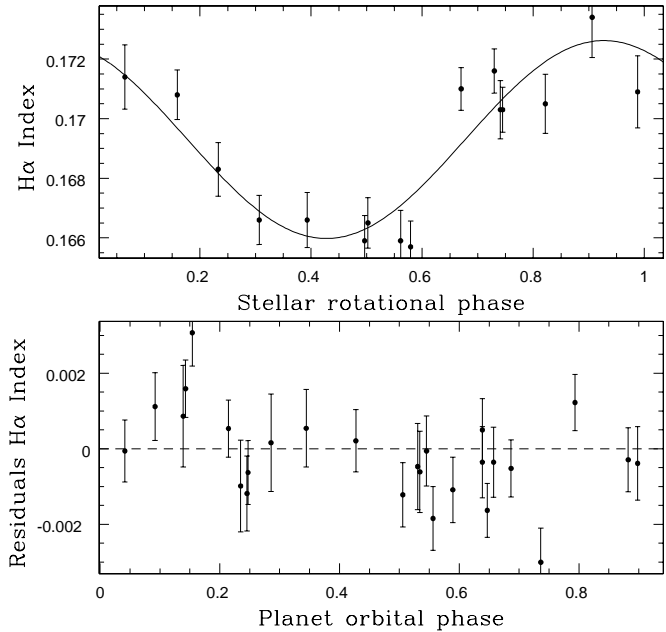}
       \caption{ Nightly averaged H$\alpha$ index as a function of rotation phase fitted by a sinusoidal function (\textit{top panel}). Residuals from the individual H$\alpha$ index after this correction are plotted as a function of the planet's orbital phase (\textit{bottom panel}). In spite of a large dispersion due to the short-term activity of the star, no specific correlation with the planet motion is observed. }
             \label{myfigH}
   \end{figure}

Shkolnik et al. (2007) identified the presence of an additional variability or flaring in the \ion {Ca} {II} cores of HD~189733,
potentially induced by interactions with the hot Jupiter. We also used the H$\alpha$ activity index to study the chromospheric variability as a function of the planet's orbit, since it is less affected by noise than the \ion {Ca} {II} index in our data. Removal of the main modulation induced by  stellar rotation was completed by estimating a sinusoidal fit of the nightly averaged H$\alpha$ indices. We did not detect significant modulation in the residual activity index with the orbital phase, as shown in Fig.~\ref{myfigH}. This result differs from the tentative conclusion by Shkolnik et al. (2007), which was based on a 4-night data set. It is possible that \ion {Ca} {II} and H$\alpha$ indices are not strongly correlated in this star, which may explain the different behavior that we observe. Alternatively, it could be another demonstration of the on/off nature of star-planet interactions (as argued by Shkolnik et al. 2007).

%
%______________________________________________________________

\section{Discussion}
 %
 
%______________________________________________________________
%                                     Two column figure (place early!)
%______________________________________________ Gamma_1 (lg rho, lg e)
   \begin{figure*}
   \centering
   \includegraphics[width=18cm]{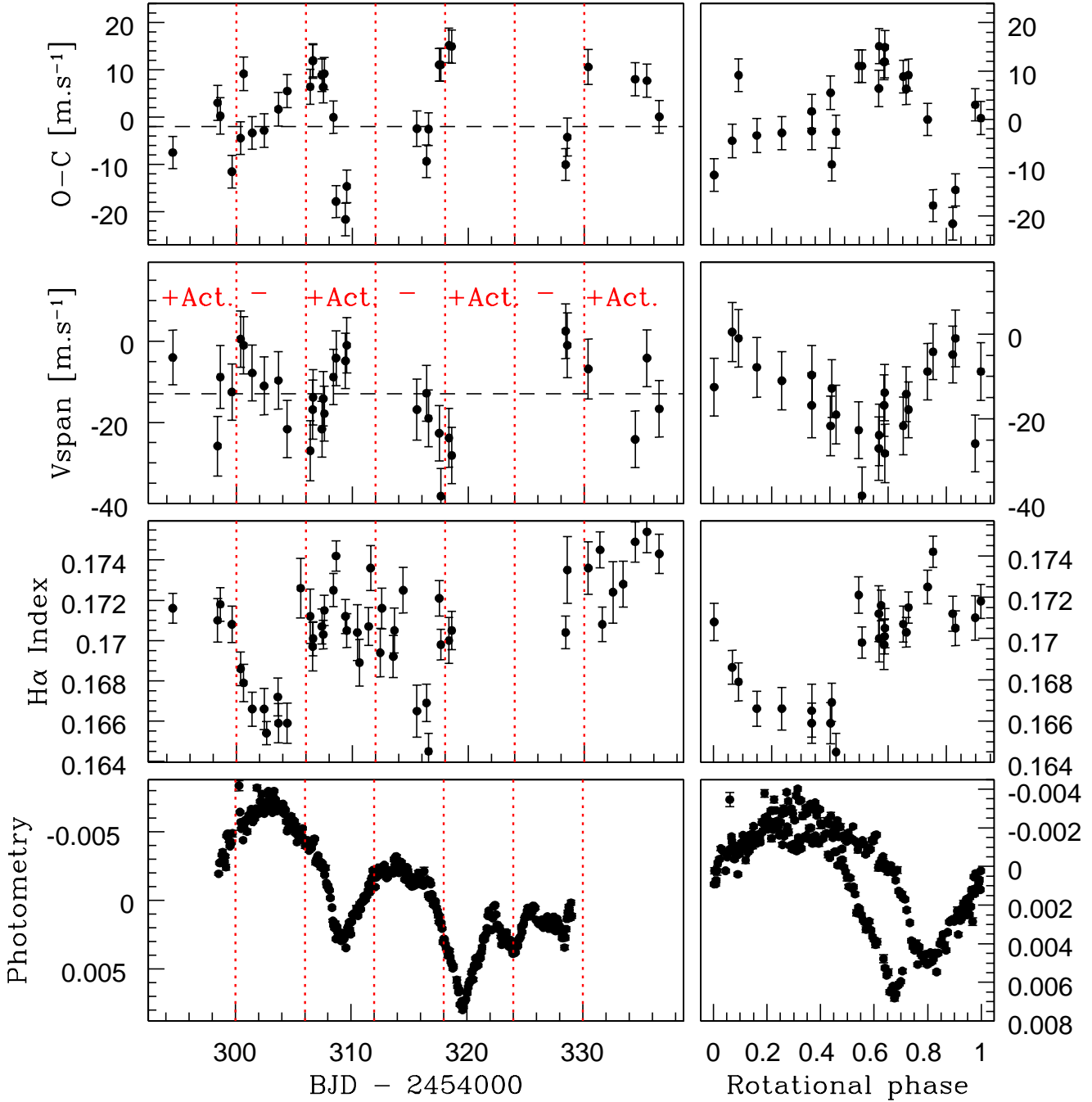}
   \caption{ {Left panels:} Residuals from the Keplerian fit (top), bisector span of the CCF profile (second panel), activity index measured in the H$\alpha$ line (third panel) and {\it MOST} photometric data (bottom) as a function of time. {Right panels:} Same parameters as a function of the stellar rotational phase, excluding data at BJD-2454000 $ \geqslant 320$ days. {In the right panel, a linear trend is substracted from the {\it MOST} data (see text).} The $T_{0}$ chosen for the rotational phase is the transit epoch of Winn et al. (2007): $T_{T}=2453988.8034$ days and $P_{rot} = 11.953$d. The unit of the {\it MOST} data is instrumental magnitude. 
V$_{span}$ error bars correspond to two times the RV error bars. The red dashed lines are based on the minimum activity index in H$\alpha$ line separated by 6 days, approximately half of the rotation period of the star. The plus (minus) signs indicate when the star is at its maximum (minimum) level of activity. 
 Periodic variations can be seen for all parameters. (O-C) and V$_{span}$ are anticorrelated (Fig.~\ref{myfig5}) and the activity index variation is quater-phase shifted compared to the RV as expected in case of spot-induced RV variations (Fig.~\ref{myfigall}). {The photometry is also in line with the activity interpretation as the star becomes darker (more spots visible) during the phases of higher activity.} }
              \label{myfig3}
    \end{figure*}

 \subsection{Periodicity of the parameters}
The period derived from an active region is not expected to be exactly the rotation period of the star. As for the Sun, HD~189733 exhibits differential rotation, with the poles rotating more slowly than the equator (Moutou et al. 2007, Far\`es et al., in prep.). We had insuficient data to determine exactly the period of our indices of activity. So, we were unable to estimate the latitude of the main active region. The rotation period of the star has been most reliably determined by Henry \& Winn (2008). They measured $P_{rot} = 11.953 \pm 0.009$ days by analysing the quasiperiodic photometric variability observed during about one year. 
In Fig.~\ref{myfig3}, both the residual RV jitter and the bisector span
appear to exhibit periodic variations on a timescale to the rotation period of the star of 12 days. This periodicity was consistent with consideration of the parameters as a function of stellar rotational phase. We excluded the data points after $BJD-2454000 = 320$ days because data points in this time interval were not in phase with the previous ones. We may infer that the main active region moved at this period.  {This is in agreement with the stellar classification for which spots are known to evolve quickly.} In the same Fig.~\ref{myfig3}, the lowest levels of activity in the H$\alpha$ line are separated by about 12 days. The 12-day variation exhibited by the H$\alpha$ index becomes more obvious when plotted as a function of the rotational phase. We excluded the data points after $BJD-2454000 = 320$ days that had clearly a higher level of activity than for the days before. The spectroscopic line profile, the indices of chromospheric emission, and the radial-velocities (O-C) therefore exhibit periodic variability around the rotation period of the star. \\

\subsection{Correlations of the parameters}
 By searching for the variations in the different parameters together in Fig.~\ref{myfig3}, we can conclude that they are consistent
 with one main active region on the stellar surface, as a first approximation for multiple active regions. Henry \& Winn (2008) estimated a lower limit of i$_{\ast}$ $\ga$ 54$^{\degr}$ for the inclination of the stellar equator with respect to the sky plane. This is consistent with the fact that the timescale of the maximum-minimum activity is around half a
rotation period.  

\begin{figure}
   \centering
   \includegraphics[width=9cm]{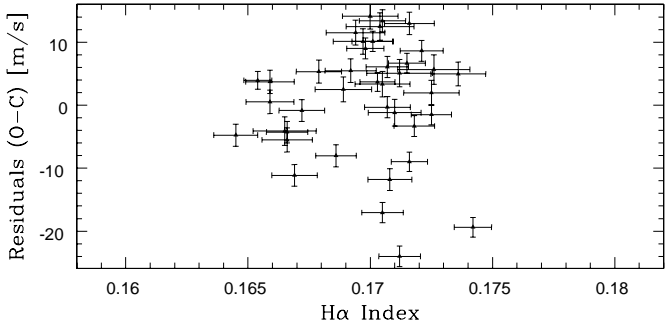}
      \caption{Radial velocities of HD189733, corrected for the signature of its known planetary companion as a
function of the H$\alpha$ spectral index. The loop pattern is a signature of a stellar surface with a main active region in rotation.
}
         \label{myfigall}
   \end{figure}

We selected the H$\alpha$ index as an indicator of activity variability since it was the most sensitive index. The activity variations exhibit roughly a quarter-phase shift with respect to the radial-velocities and line profile
variations. This fact is in agreement with an observed stellar activity \citep{queloz} and a model of a dark spot rotating with the stellar surface and moving in and out of view. In Figure~\ref{myfigall}, we also plot the residuals from the Keplerian fit (O-C) as a function of the
chromospheric activity index. We observe a loop pattern as expected for an active region, which features a dark spot rotating on the stellar surface as already described by Bonfils et al. (2007). The quarter-phase shift and the loop pattern results from the same correlation. When the activity level is at its maximum and the spot is on the central meridian of the star, it occupies symmetrically the approaching and the receding halves of the stellar disk, and we do not expect the RV to be perturbed. Similarly, when the spot is out of sight, the activity level is at a minimum and the impact on the RV should be negligible, in addition to the impact on the line shape. On the other hand, when the spot appears or disappears on the visible stellar disk, it lies on the rotationally blue-shifted or red-shifted half part of the stellar surface and induces an asymmetry in the velocity distribution of the emitted flux. Thus, when the dark spot appears (disappears), the activity index is at its middle level and the RV effect is minimum (maximum), while the V$_{span}$ is anticorrelated with the RV.

%                                                One column figure
%----------------------------------------------------------- S_vib
%
%______________________________________________________________

 \subsection{ {\it MOST} photometric data}
Our observations of HD 189733 can be compared with those taken simultaneously 
with the {\it MOST} microsatellite. {\it MOST} (Walker et al. 2003; Matthews et al. 2004) is 
an optical microsatellite that is designed to observe nearby, relatively 
bright stars through a single broadband (350-700 nm) filter using a 
15/17.3 cm Rumak-Maksutov telescope.
{\it MOST} observed HD 189733 for over $\sim$31 days in 2007. The 2007  observations, reduction and resulting analysis is discussed in Rowe et al. (in prep.). For comparison with the {\it SOPHIE} observations, we present the {\it 
MOST} observations of HD 189733 in 2007 in {Fig.~\ref{myfig3}} (bottom panels). The {\it 
MOST} lightcurve has the transits of the known planet removed, the data is 
binned to the orbital period of the satellite ($P$ $\sim$ 101.43 minutes). {A 
long-term linear trend of 5.7$\times$10$^{-4}$ mag day$^{-1}$ is removed in the right panel of the Fig.~\ref{myfig3} in order to emphazise the correlations with the other parameters.} The comparison stars do not exhibit this drift, suggesting that the drift is real.  Since
the timescale for the drift appears to be much longer than the length
of the observations, one can only speculate to its cause.
As can be seen in the bottom panel of 
Fig.~\ref{myfig3}, HD 189733 displayed $\sim$1\% variations in flux that were {most likely} due to starspot 
modulation over $\sim$2.5 rotation cycles of the star in optical light  
during the {\it MOST} observations. In 2006, during the same period,  {\it MOST} observations of HD 189733 displayed $\sim$3\% variations in flux (Croll et al. 2007). The difference in the amplitude of the variation may be due to a decrease in amplitude of the stellar activity jitter with time due to a long-term activity cycle of the star. This interpretation is only speculation but would agree with the decrease in the (O-C) residuals around the fit in RV observed by the successive RVs campaigns in 2005, 2006, and 2007 discussed in Sect.~4.

{\it MOST} photometric data exhibit variations in the stellar flux with a periodicity of about {11} days. When the MOST data were acquired simultaneously with the SOPHIE data, the maximum flux agrees with the minimum of activity, showing that the variation in emitted flux of the star is dominated by dark spots. {Between BJD-2454000 = 318 and 329 days}, the photometric curve seems to present a new main active region. The minimum flux at BJD-2454000 = 319.5 days comes just 10 days after the previous minimum and is also deeper than the previous one. Moreover, a third shallow minimum occurs at BJD-2454000 = 324 days, and the maximum flux at BJD-2454000 = 329 days comes later (12 days) than the previous maximum. This corresponds to the fact that the radial-velocities residuals, V$_{span}$, and H$\alpha$ index data after BJD-2454000 = 328 days are not in phase with their previous values before BJD-2454000 = 320 days. 

%
%______________________________________________________________

\subsection{Profile of the CCF}
 
\begin{figure}[top]
   \centering
   \includegraphics[width=9cm]{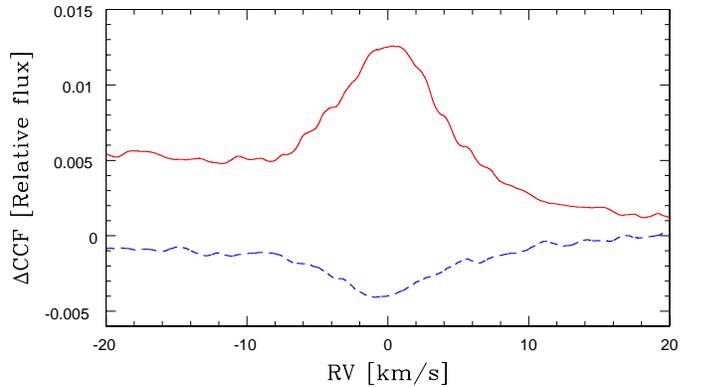}
      \caption{ Profile of the difference between a mean CCF computed at maximum activity level and a mean CCF of all the data available (solid line) and the difference between a mean CCF computed at minimum activity level and a mean CCF of all the data available (dashed line). This illustrates the behavior of the spectral lines with stellar activity.
              }
         \label{myfig8}
   \end{figure}

Using the fact that the CCF is equivalent to an average of all spectral lines, we compared the profile of the CCF when the star was both at its maximum level of activity and its lowest activity level. The abscissa of each CCF was first corrected for the orbit of the planet. All the available spectra unaffected by the background Moon light were used to compute the mean CCF. On the other hand, we summed selected spectra to compute a mean CCF at maximum activity level (MaxCCF) and a mean CCF at minimum activity level (MinCCF). The mean CCF was subtracted from MaxCCF and MinCCF. The resulting profiles are plotted in Fig.~\ref{myfig8}. When the star was at its maximum level of activity, the difference (MaxCCF-Mean) was positive whereas at its minimum level of activity, the difference (MinCCF-Mean) was negative. This result could be explained by the presence of dark spots when the star was at its maximum level of activity. Dark spots have a lower temperature and exhibit different absorption lines than those of the star. The spectrum of the dark spot may be closer to that of an M-dwarf. When there are dark spots on the stellar surface, the lines of the K2V star are shallower. This result shows that the parameters of the CCF may be used to characterize the variation in the stellar activity.  
%
%______________________________________________________________
 %
%                                                One column figure
%----------------------------------------------------------- S_vib

\subsection{Improvement of RV measurements}

In a similar way to Melo et al. (2007), we corrected the radial-velocity measurements for the stellar activity and refined the orbital parameters of the planetary system. We used the least-square fit  between the residuals (O-C) and V$_{span}$ according to Fig.~\ref{myfig5}, as determined in Sect.~4.  We added the least-square fit between (O-C) and V$_{span}$ to the residuals (O-C) data points. We then adjusted the most accurately determined Keplerian fit to the corrected radial velocities (new residuals summed to values of the previous orbit). As in Sect.~3, we fixed the orbital period, the time of periastron, and the eccentricity. We derived lower residuals from the fit of 3.7 m s$^{-1}$, compared with 9.1 m s$^{-1}$ before correction. These lower residuals are in agreement with the estimated current systematics on {\it SOPHIE}  measurements discussed in Sect.~2. The error bars on the semi-amplitude of the orbit are also smaller with $K$= 200.6 $\pm$ 0.9 m s$^{-1}$, compared with $K$= 200.6 $\pm$ 2.3 m s$^{-1}$ before correction. Using the mass of the star $M_{\star}$= 0.82 $\pm$ 0.03 M$_{\sun}$ given by Bouchy et al. (2005),  we found a planet mass of $M_{P}$= 1.13 $\pm$ 0.03 M$_{Jup}$. Our result agrees exactly with the value of 1.13 $\pm$ 0.03 M$_{Jup}$ reported by Winn et al. (2006), and it is also in agreement with the Bouchy et al. (2005) value of 1.15 $\pm$ 0.04 M$_{Jup}$. The current uncertainty in $M_{P}$ is dominated by the error in $M_{\ast}$.

%
%______________________________________________________________ 

\section{Conclusion}
We have observed the transiting planetary system HD 189733 with the high-resolution {\it SOPHIE}  spectrograph, with the aim of correcting the parameters of the planetary orbit for the effect of radial-velocity variations due to activity on the stellar surface. We refined the HD~189733b orbital parameters and placed limits on both the orbital eccentricity and long-period variation. After subtracting the orbital motion of the known planet, the radial-velocity residuals were correlated with the spectroscopic activity indices and the shape of spectral lines. 
Chromospheric activity tracers and surface photometry were modulated mainly by the 12-day rotational period of the star. However, the relation that we have derived between the bisector span and radial-velocity residuals is the most accurate yet and allows a precise correction for stellar jitter. The error in the derived semi-amplitude of the orbit is reduced by a factor of 2.5 after applying this correction law and reduced to close to the instrumental limit, allowing a more precise estimate of the companion's mass.   

This result is promising for radial-velocity surveys, which cannot continue to avoid targeting active stars. Additionally, radial-velocity instruments have to perform follow-up observations for photometric planet search programs, which are less sensitive to the stellar activity for detecting transits. In case the use of a single correlation is insufficient, the use of several indices allows one to be more confident of accounting for the main active regions and provides stellar-variability diagnostics. In the present study, the bisector span was found to be the most reliable parameter. In the case of the bright star HD 189733, this method of correction is a good compromise. But, if the correction decreases the activity jitter, it also increases the V$_{span}$ photon-shot noise in the corrected radial-velocity measurements. This method has therefore to be tested further on other systems, to estimate the confidence in the corrected RVs and, subsequently, to apply similar analyses in planet-search programs. Finally, a clearer understanding of the stellar intrinsic variability will have a positive impact on identifying and studying activity features potentially triggered by star-planet interactions.

%
%---------------------------------------------------------------------------------

\begin{acknowledgements}
   N.C.S. would like to thank the support from Funda\c{c}\~ao para a 
Ci\^encia e a Tecnologia, Portugal, in the form of a grant (references 
POCI/CTE-AST/56453/2004 and PPCDT/CTE-AST/56453/2004), and
through programme Ci\^encia\,2007 (C2007-CAUP-FCT/136/2006). 
   We would like to thank the referee, M. K\"urster, for all his useful comments, which helped to improve this paper. 
\end{acknowledgements}

%
%_______________________________________________

\end{document}